\documentclass{article}
\usepackage[left]{lineno}

\usepackage{comment}
\renewcommand\appendix{\par
  \setcounter{section}{0}
  \setcounter{subsection}{0}
  \setcounter{figure}{0}
  \setcounter{table}{0}
  \renewcommand\thesection{Appendix \Alph{section}}
  \renewcommand\thefigure{\Alph{section}\arabic{figure}}
  \renewcommand\thetable{\Alph{section}\arabic{table}}
}
\usepackage[normalem]{ulem}
\usepackage{float}
\usepackage{graphicx}
\usepackage{epstopdf }
\usepackage{caption}
\usepackage{subcaption}
\usepackage{here}
\usepackage{amsmath,amssymb,amsthm}
\usepackage{color}
\theoremstyle{plain}

\usepackage{multirow}
\theoremstyle{definition}

\usepackage{soul}
\usepackage{chngcntr}
\usepackage{listings}

\usepackage{soul}
\newcommand{\mv}[1]{{\boldsymbol{#1}}}

\usepackage{amsopn}
\usepackage{amsfonts}
\usepackage[mathscr]{eucal}
\usepackage{fancybox}
\usepackage{bm} %bold Greek font
\usepackage[subnum]{cases}
\usepackage[normalem]{ulem}
\usepackage{marginnote} %Package helping with margin notes
\title{Online estimation of driving events and fatigue damage on vehicles}
\author{Roza Maghsood and Jonas Wallin\\
Mathematical Sciences,\\ Chalmers University of Technology,\\ SE-412 96 G\"oteborg, Sweden.}
\date{}
\begin{document}
\maketitle
%%%%%%%%%%%%%%%%%%%%%%%%%%
\begin{abstract}
Driving events, such as maneuvers at slow speed and turns, are important for durability assessments of vehicle components. By counting the number of driving events, one can estimate the fatigue damage caused by the same kind of events. Through knowledge of the distribution of driving events for a group of customers, the vehicles producers can tailor the design, of vehicles, for the group. In this article, we propose an algorithm that can be applied on-board a vehicle to online estimate the expected number of driving events occurring, and thus be used to estimate the distribution of driving events for a certain group of customers. Since the driving events are not observed directly, the algorithm uses a hidden Markov model to extract the events. The parameters of the HMM are estimated using an online EM algorithm.  The introduction of the online EM is crucial for practical usage, on-board vehicles, due to that its complexity of an iteration is fixed. Typically, the EM algorithm is used to find the, fixed, parameters that maximizes the likelihood. By introducing a fixed forgetting factor in the online EM, an adaptive algorithm is acquired. This is important in practice since the driving conditions changes over time and a single trip can contain different road types such as city and highway, making the assumption of fixed parameters unrealistic.  Finally, we also derive a method to online compute the expected damage.

%By measuring the loads acting on components, we can clarify which occasions will generate high forces to determine the fatigue damage in service. Different methods have been used to detect driving events such as hidden Markov models (HMMS). The HMMs are reliable and robust methods for identification and they are very useful for event recognition in signal processing. We focused on identifying the driving events using online HMMs. The method can update the HMM parameters recursively over time by updating the measurements. Online or recursive HMM is useful because we can use the new measurements over the time without storing the previous observations. In this article, we use an EM-online algorithm to estimate the HMM parameters. The estimation results shows that online HMMs works as well as general EM algorithm.
\end{abstract}

\vspace{5mm}
\section{Introduction}
%\linenumbers
When designing vehicles components it is important to know the distributions of loads expected to act on them. The life time of a component in a vehicle--  such as control arms, ball joints, etc.-- is determined by its strength and the loads acting on it.
Where the effect of a given force acting on a component is well known, the distributions of loads, and hence forces,
are more random. This is because the distribution of the loads depends on the driving environment, driver's behavior, usage of the vehicle, and other things. For a more detailed description of loads acting on vehicles see \cite{Johannesson2}.
 
 Although it is not financially possible to design a vehicle for specific customer, it is important to 
 tailor the design for groups of customers, depending on, for instance, geographical regions and usage. Obviously, components weakly designed for the specific environments  leads to increased costs due to call-backs and badwill for the company, while too heavily designed components gives increased material cost and unnecessarily heavy vehicles. 
 
Traditionally, one has used a specially equipped test vehicle to study the distributions of customer loads. This gives very precise measurements, but with disadvantage of a statistically small sample size for the studied group. In addition, it is a very expensive way of acquiring data. However, all modern vehicles are equipped with computers measuring many signals, known as Controller Area Network (CAN) bus data, where the signal is for instance speed and lateral acceleration. The goal of this article is to develop a statistical algorithm that uses these signals, to extract information about the driving events for the specific vehicle. This data can then be collected from several vehicles to generate a load distribution for groups of customers.

The desired algorithm needs several key properties to be practically useful:
First, it obviously needs to be able to extract the driving events from the CAN data.
Second, since the data will be extracted over long periods of time the computational cost of estimation of the driving events needs to be low. It is also desirable that the method does not require the storage of all the data. 
Finally, the algorithm should allow for changing frequency of driving events over time, since the frequency of driving events changes depending on the driving environment such as highway driving or city driving.

To address the first property, our algorithm uses a hidden Markov Model (HMM) to extract the driving events from the CAN data. More specifically each state in the HMM represents a driving state  where we define a driving event  as a sequence of consecutive driving states. The CAN data for a given driving state is assumed to follow a generalized Laplace (GAL) distribution. Laplace distributions are well known methods to describe responses measured on driving vehicles, see \cite{Rychlik13}, \cite{Kvanstrom} and \cite{Bogsjo}.  The idea of using HMMs to identify driving events has previously been used in for example Maghsood and Johannesson \cite{Maghsood2, Maghsood1}, Mitrovi\'c \cite{Mitrovic1,Mitrovic2} and Berndt and Dietmayer \cite{Berndt}.

For the HMM we divide the parameter into two sets: the transition matrix, which is vehicle type independent, depending rather on the driving environment, the driver's behavior etc. The parameter of the GAL distribution is vehicle type specific, and can thus be found in laboratory tests or in proving grounds. Thus the second property, in the case of an HMM, is equivalent to efficiently estimating the transition matrix of driving states. In previous articles, the EM algorithm has been used successfully to estimate the transition matrix,\cite{Maghsood3}; however an iteration of the algorithm has computational complexity  $\mathcal{O}(n)$ (where $n$ is the number of observation) and is thus not practically feasible. Here, we instead propose using the online EM algorithm from Capp\'e \cite{Cappe1} to estimate the matrix. This gives the desired computational efficiency, since one iteration of the algorithm has a computational cost of $\mathcal{O}(1)$.

The final property is addressed by using a fixed forgetting factor in the online EM algorithm. Capp\'e \cite{Cappe1} proposes a diminishing forgetting factor to ensure that the EM algorithm converges to a stationary point. However, this is not the goal here and we do not want the algorithm to converge to a stationary point but rather be an adaptive algorithm. The usage forgetting factors is a well-studied area in automatic control, time series analysis and vehicle engineering  \cite{arvastson2000asymptotic, ljung1983theory,vahidi2005recursive}.

Further, the algorithm also calculates online the expected damage for a given component. This could be useful for the specific vehicle, on which the algorithm is applied, by using the expected damage to tailor service times to specific vehicle and components. 

The paper is organized as follows: In the second section, the HMM and the proposed online algorithm are presented. In the third section, the method for estimating the fatigue damage is proposed. In the forth section, the algorithm is applied to simulated data to verify its performance, and it is also evaluated on real data, CAN data from a Volvo truck. The final section contains the conclusions of the paper.

\section{Hidden Markov models}
\label{sec:HMM}
%%\mn{JW: The first sentence is waaay to general. What does different pracital situation mean? Cite rater the use in speech processing: article , The application of hidden Markov models in speech recognition
%and for finance (I think ): Finite mixture and Markov switching models. But also cite Cappe }
Hidden Markov models are statistical models often used in signal processing, such as speech recognition and modeling the financial time series, see for instance Capp\'e \cite{Cappe} and Frühwirth-Schnatter \cite{F-Schnatter}. An HMM is a bivariate Markov process $\left\{Z_t,Y_t\right\}^{\infty}_{t=0}$ where the underlying process $Z_t$ is an unobservable Markov chain and is observed only through the $Y_t$. The observation sequence $Y_t$ given $Z_t$ is a sequence of independent random variables and the conditional distribution of $Y_t$ depends only on $Z_t$.

%The model consists of two processes $\left\{Z_t\right\}^{\infty}_{t=0}$ and $\left\{Y_t\right\}^{\infty}_{t=0}$ where $X_t$ is a hidden Markov chain and $Y_t$ is the observations. The observed process $Y_t$ given $X_t$ is a sequence of independent random variables and the conditional distribution of $Y_t$ depends only on $Z_t$.

In this article, all HMMs  are such that $Z_t$ takes values on a discrete space $\{1,2,\ldots,m\}$, and the HMM is determined by two sets of parameters. The first set is the transition probabilities of Markov chain $Z_t$:
\begin{equation}\label{Tra}
q(i,j)=P(Z_{t+1}=j|Z_{t}=i),\ i,j=1,2,...,m.
\end{equation}
 The second set is the parameter vector, $\mv{\theta}$, of the conditional distribution of $Y_t$ given $Z_t$:
\begin{equation}\label{Emi}
g_{\mv{\theta}}(i,y_t)=f_{Y_t}(y_t|Z_{t}=i;\mv{\theta}), \ i=1,2,...,m, \ y_t\in \mathbb{R}.
\end{equation}
 
\noindent
 Here, we denote the set of parameters by $\Theta = (\mv{Q},\mv{\theta})$ where $\mv{Q}=(q(i,j))$ for $ i,j=1,2,...,m$.

In an HMM, the state where the hidden process will start is modeled by the initial state probabilities $\mv{\pi}=(\pi_{i})$, where $\pi_{i}$ is denoted by:
\begin{align*}
\pi_{i}=P(Z_{0}=i), \ i=1,2,...,m 
\end{align*}
with $\sum^{m}_{i=1}\pi_{i}=1$.

\subsection{Parameter estimation}
For the parameter estimation in this article we use the EM (expectation maximization) algorithm, which is described below. The principle aim is to estimate  the transition matrix $\mv{Q}$ based on an observation sequence. For this, we use an online EM algorithm, derived in \cite{Cappe1}. To introduce the algorithm we first describe the EM algorithm and then describe the modification needed for online usage of the algorithm.

 In our study, the parameter $\mv{\theta}$ is not estimated recursively, but rather found through maximum likelihood estimation
 on a training set. This is because the conditional distribution of $Y_t$ given $Z_t$ in our case study represents the vehicle specific data which can be estimated under well-defined conditions on the proving ground.

%The reason for this is that the conditional distribution of observations given certain hidden state, describes the property of driving events which is constant over time. But, the differences between types of roads can affect the transition probabilities and a transition matrix describes the duration of the events. Therefore, it could be reasonable to update the transition matrix over time. 

\subsection{The EM algorithm}
Here, we present the EM algorithm following Capp\'e \cite{Cappe1}. The EM algorithm is a common method for estimating the parameters in HMMs. It is an optimization algorithm to find the parameters that maximize the likelihood. The algorithm is both robust -- it does not diverge easily-- and is often easy to implementation. 

The EM algorithm is an iterative procedure. If the distribution of complete-data $(Z_t,Y_t)$ given $Z_{t-1}$, $p(z_t,y_t|z_{t-1})$, belongs to an exponential family, then the $n^{th}$ iteration consists of the two following steps:
\begin{itemize}
\item The E-step, where the conditional expectation of the complete-data sufficient statistics, $s(Z_{t-1},Z_t,Y_t)$, given the observation sequence $Y_{0},Y_{1},...,Y_{t}$ and $\Theta^{(n)}$, is computed, 
\begin{align}\label{Suff}
\mv{S}^{(n+1)}_t=\frac{1}{t}E\left[\sum^{t}_{l=1}s(Z_{l-1},Z_l,Y_l)\bigg|Y_0,...,Y_t;\Theta^{(n)}\right],
\end{align}
\item The M-step, where the new parameter value $\Theta^{(n+1)}$ is calculated using $\mv{S}^{(n+1)}_t$, which can be formulated as $\Theta^{(n+1)} = f(\mv{S}^{(n+1)}_t)$.
\end{itemize}
The sequence $\Theta^{(n)}$ converges to a stationary point of the likelihood function, for more details see \cite{Cappe1}. 

%It should be noted that a sufficient statistic is a function of observations, which contains all information about the parameters in the data.  If the distributions of $Y_t,X_{t},$ given $X_{t-1}$ belongs to an exponential family, there exists a sufficient statistic of the form $s(X_{t-1},X_t,Y_t)$.

%In this article, the parameter of interest is the transition matrix $\mv{Q}=(q(i,j))$ where the state transition probabilities $q(i,j)$ are defined by~$(\ref{Tra})$. Then, the E-step consists of computing
For our specific model, where the parameter of interest is $\mv{Q}$, the sufficient statistics in the E-step is:
\begin{align}\label{Suff1}
S^{(n+1)}_t(i,j)=\frac{1}{t}E\left[\sum^{t}_{l=1} I (Z_{l-1}=i,Z_{l}=j)\bigg|Y_0,...,Y_t;\Theta^{(n)}\right].
%&S^{g}_{t}(i)=
%\begin{cases}
%\frac{1}{n}E[\sum^{t}_{l=0} I (X_{l}=i,Y_l=y)|Y_0,...,Y_t], &\mbox{if } Y_t \ \mbox{is discrete},\\
%\frac{1}{n}E[\sum^{t}_{l=0} I (X_{l}=i)s(Y_l)|Y_0,...,Y_t], &\mbox{if } Y_t \ \mbox{is continuous}.
%\end{cases}
\end{align}
Thus $S_t(i,j)$ is the expected number of transitions from state $i$ to state $j$ given $Y_0,...,Y_t$ and $\Theta$. For $\mv{Q}=(q(i,j))$, the M-step is given by:
\begin{align}\label{ReEstiTra}
q^{(n+1)}(i,j)= \frac{S^{(n+1)}_{t}(i,j)}{\sum^{m}_{j=1}S^{(n+1)}_{t}(i,j) }.
\end{align}
\subsubsection{Recursive formulation of the E-step}
Zeitouni and Dembo \cite{Zeitouni} noted that the conditional expectation of the complete-data sufficient statistics $\mv{S}_t$ can be computed recursively.  To see this, define
\begin{align}\label{Recursion1}
&\phi_{t}(k)=P(Z_t=k|Y_0,...,Y_t;\Theta),\\
&\rho_{t}(i,j,k)=\frac{1}{t}E[\sum^{t}_{l=1} I (Z_{l-1}=i,Z_{l}=j)|Y_0,...,Y_t, Z_t=k ;\Theta],
\end{align}
then $S_{t}(i,j)$ can be written as $S_{t}(i,j)=\sum^{m}_{k=1} \phi_{t}(k)\rho_{t}(i,j,k)$.\\

\noindent
Note that $(\mv{\phi}_t)_k=\phi_{t}(k)$ is an $N$-dimensional (row) vector. For a vector $\mv{a}$, let $\mv{D}(\mv{a})$ be a diagonal matrix where $\mv{D}(\mv{a})_{kk} = a_k$. The recursive implementation of the EM algorithm, using the observation sequence $Y_{0},Y_{1},...,Y_{T}$,  is initialized with 
$$
\mv{\phi}_{0}=\frac{\mv{\pi}\mv{D}(g_{\mv{\theta}}(k,y_0))}{(\mv{\pi}\mv{D}(g_{\mv{\theta}}(k,y_0)))\mv{1'}}, \ \text{and} \
\rho_0(i,j,k)=0,
$$
for all $1 \le i,j,k \le m$. Let $g_{\mv{\theta}}(y_t)=\left(g_{\mv{\theta}}(1,y_t), g_{\mv{\theta}}(2,y_t), ..., g_{\mv{\theta}}(m,y_t)   \right)$ and $\mv{1}=\left(1,1,...,1 \right)$. Then, for $n^{th}$ iteration and $t\geq1$, the components are updated as follows:
\begin{eqnarray}
\mv{\phi}_{t+1}&=&\frac{\mv{1}(\mv{D}(\mv{\phi}_{t})\mv{Q}^{(n)}\mv{D}(g_{\mv{\theta}}(y_t)))}{\mv{1}(\mv{D}(\mv{\phi}_{t})\mv{Q}^{(n)}\mv{D}(g_{\mv{\theta}}(y_t)))\mv{1'}},\label{Recursion2} \\ 
\rho_{t+1}(i,j,k)&=&\gamma_{t+1}I(j-k)r_{t+1}(i|j)+(1-\gamma_{t+1})\sum^{m}_{k'=1}\rho_{t}(i,j,k')r_{t+1}(k'|k), \label{Recursion3}
\end{eqnarray}
where $\mv{r}_{t+1}=\mv{D}(\mv{\phi_{t}}./\mv{1}(\mv{D}(\mv{\phi}_{t})\mv{Q}^{(n)}))\mv{Q}^{(n)}$ and $./$ represents the element-wise division of two matrices. The forgetting factor, $\gamma_{t}$, equals $1/t$.\\
%$\mv{D}(g_{\theta}(i,y_t))$ is a diagonal matrix which its  $i^{th}$ element is $g_{\theta}(i,y_t)$ given by~$(\ref{Emi})$, 

%%%%%%%%%%%%%%%%%%%%%%%%%%%
%The recursive implementation of the EM algorithm, using the observation sequence $Y_{0},Y_{2},...,Y_{T}$, will start by two initial quantities $\phi_{0}(k)=\frac{\pi_{k}g_{\theta}(k,y_0)}{\sum^{N}_{k'=1}\pi_{k'}g_{\theta}(k',y_0)}$ and $\rho_0(i,j,k)=0$ for all $1 \le i,j,k \le N$. Which is updated recursively by:
%\begin{eqnarray*}
%\phi_{t+1}(k)&=&\frac{\sum^{N}_{k'=1}\phi_{t}(k')q_{t}(k',k)g_{\theta}(k,y_{t+1})}{\sum^{N}_{k',k''=1}\phi_{t}(k')q_{t}(k',k'')g_{\theta}(k'',y_{t+1})},\label{Recursion2} \\ 
%\rho_{t+1}(i,j,k)&=&\gamma_{t+1}I(j-k)r_{t+1}(i|j)+(1-\gamma_{t+1})\sum^{N}_{k'=1}\rho_{t}(i,j,k')r_{t+1}(k'|k), \label{Recursion3}
%\end{eqnarray*}
%where $r_{t+1}(i|j)=\frac{\phi_{t}(i)q_{t}(i,j)}{\sum^{N}_{i'=1}\phi_{t}(i')q_{t}(i',j)}$, and $\gamma_{t}=1/t$ is a forgetting factor.\\
%%%%%%%%%%%%%%%%%%%%%%%%%%%%

Note that in $n^{th}$ iteration of EM algorithm, all elements in $\phi_1, \phi_2, ..., \phi_t$ and $\rho_1, \rho_2, ..., \rho_t $ depends on $\mv{Q}^{(n)}$.  Thus, for updating $\mv{Q}$ in $(n+1)^{th}$ iteration, all elements of the two quantities need to be recalculated. Therefore one needs to store the entire observation vector to use the EM-algorithm.

\subsection{Online estimation of HMM parameters}
\label{sec:Q}
%In an online EM algorithm, the parameters are estimated as a new observation arrives without needing to store the previous observations.

As we will see soon, the online EM algorithm remedies the issue of requiring the entire observation vector to estimate parameters.  Here we use the notation $\hat{\mv{Q}}_t$ rather then $\mv{Q}^{(t)}$. This is because, as we will see, one can not compute more than one iteration at each time point $t$ for the online EM.

The terms $\hat{\mv{\phi}}_{0}$ and  $\hat \rho_0(i,j,k)$ are initialized the same way as in the regular EM algorithm.
For $t=0, 1,\ldots$ the components are updated as follows: (the E-step)
\begin{eqnarray}
\hat {\mv{\phi}}_{t+1}&=&\frac{\mv{1}(\mv{D}(\mv{\phi}_{t})\hat{\mv{Q}}_t\mv{D}(g_{\mv{\theta}}(y_t)))}{\mv{1}(\mv{D}(\mv{\phi}_{t})\hat{\mv{Q}}_t\mv{D}(g_{\mv{\theta}}(y_t)))\mv{1'}},\label{Recursion4} \\ 
\label{eq:rhohat}
\hat \rho_{t+1}(i,j,k)&=&\gamma_{t+1}I(j-k)\hat r_{t+1}(i|j)+(1-\gamma_{t+1})\sum^{m}_{k'=1}\hat \rho_{t}(i,j,k')\hat r_{t+1}(k'|k), \label{Recursion5}
\end{eqnarray}
where $\hat{\mv{r}}_{t+1}=\mv{D}(\hat{\mv{\phi}_{t}}./\mv{1}(\mv{D}(\hat{\mv{\phi}}_{t})\hat{\mv{Q}}_t))\hat{\mv{Q}}_t$.\ And in the M-step, the transition matrix $\hat{\mv{Q}}_{t+1}=(\hat{q}_{t+1}(i,j))$ is updated by:
\begin{align}\label{onlineEstiTra}
\hat{q}_{t+1} (i,j)=\frac{\hat S_{t+1}(i,j)}{\sum^{m}_{j=1}\hat S_{t+1}(i,j) },
\end{align}
where $\hat S_{t+1}(i,j)=\sum^{m}_{k=1} \hat \phi_{t+1}(k)\hat \rho_{t+1}(i,j,k)$.\\

%%%%%%%%%%%%%%%%%%%%%%%%%%%%%%%%%%%%%%%%%
%The two matrices needed in the online estimation are initiated as $\hat \phi_{0}(k)=\frac{\pi_{k}g_{\theta}(k,y_0)}{\sum^{N}_{k'=1}\pi_{k'}g_{\theta}(k',y_0)}$ and $\hat \rho_0(i,j,k)=0$ for $1 \le i,j,k \le N$. And then updated  recursively, in the E-step, by:
%\mn{JW: is matrix form better here also?}
%\begin{eqnarray}
%\hat \phi_{t+1}(k)&=&\frac{\sum^{N}_{k'=1}\hat \phi_{t}(k')\hat q_{t}(k',k)g_{\theta}(k,y_{t+1})}{\sum^{N}_{k',k''=1}\hat \phi_{t}(k')\hat q_{t}(k',k'')g_{\theta}(k'',y_{t+1})},\label{Recursion4} \\ 
%\hat \rho_{t+1}(i,j,k)&=&\gamma_{t+1}I(j-k)\hat r_{t+1}(i|j)+(1-\gamma_{t+1})\sum^{N}_{k'=1}\hat \rho_{t}(i,j,k')\hat r_{t+1}(k'|k), \label{Recursion5}
%\end{eqnarray}
%where $\hat r_{t+1}(i|j)=\frac{\hat \phi_{t}(i)\hat q_{t}(i,j)}{\sum^{N}_{i'=1}\hat \phi_{t}(i')\hat q_{t}(i',j)}$. In the M-step, the state transition probability $Q^{t+1}$ is updated by:
%\begin{align}\label{onlineEstiTra}
%Q^{t+1}_{i,j} =\frac{\hat S_{t+1}(i,j)}{\sum^{N}_{j=1}\hat S_{t+1}(i,j) },
%\end{align}
%where $\hat S_{t+1}(i,j)=\sum^{N}_{k=1} \hat \phi_{t+1}(k)\hat \rho_{t+1}(i,j,k)$.\\
%%%%%%%%%%%%%%%%%%%%%%%%%%%%%%%%%%%%%%%%%%%%%%%%

As can be seen, Eqs.~$(\ref{Recursion4})$ and~$(\ref{Recursion5})$ are the modifications of Eqs.~$(\ref{Recursion2})$ and~$(\ref{Recursion3})$ where $\hat{\mv{\phi}}_{1}, \hat{\mv{\phi}}_{2}, ..., \hat{\mv{\phi}}_{t}$ and $\hat \rho_{1}, \hat \rho_{2}, ..., \hat \rho_{t}$ did not depend on the parameter $\mv{Q}$, but rather $\hat{\mv{Q}}_{t}$, and thus do not need to be recalculated. 

In the proposed online EM algorithm by Capp\'e \cite{Cappe1}, a decreasing sequence of forgetting factors  $\{\gamma_{t}\}^{\infty}_{t=1}$ is chosen such that $\sum^{\infty}_{t=1}\gamma_{t}=\infty$ and $\sum^{\infty}_{t=1}\gamma^{2}_{t}<\infty$. The choice of $\gamma_{t}$ strongly affects the convergence of the parameters. To converge to a stationary point one can choose $\gamma_{t}=1/t^{\alpha}$ with $0.5<\alpha<1$, which is the common choice suggested in \cite{Cappe1}. By setting $\gamma_{t}$ to a fixed value, the algorithm will never converge to any fixed point but behave like a stochastic processes. As we will see later, this can be useful when the data comes from a non-stationary process, where the parameters are not fixed over time.

\subsubsection{Setting forgetting factor}
When using a fixed value for $\gamma_t \, (=\gamma)$ it is crucial that this value is well chosen. A smaller $\gamma$ gives a more stable parameter trajectory, at the price of a slower adaptation. In the present form, it can be hard to see what a reasonable value of $\gamma$ is. To show this clearly, we introduce two explanatory parameters ($K$, $R$), which represent the weight, $R$, that is put on the $K$ latest observations, when estimating $\mv{Q}$. So for instance, if $K = 100$, and $R=0.9$, then the weight given to the hundred latest observations is such that, they represent $90\%$ of the information from the data used to estimate the parameters.

To link the parameters $K$ and $R$ to $\gamma$, note that (\ref{eq:rhohat}) is approximately a geometric series with ratio $\gamma$, thus approximately it holds that\begin{align}\label{gam}
\gamma  \sum_{i=0}^K (1-\gamma)^i = R.
\end{align}
This gives an explicit $\gamma$ for each $(R,K)$.

A further issue is that in general, one observations does not contain equal information about all the entires in $\mv{Q}$, some states (events) might occur rarely and thus most observations contain no information about the corresponding column in the transition matrix. To address this, one can set a separate $\gamma$ for each column. One way is to set $\gamma_{t,i} = \gamma \cdot (\mv{\pi}_t)_i$ where $\mv{\pi}_t$ is the averaged stationary distribution vector defined below.

\subsection{Online estimation of the number of events}
In previous work, see Maghsood, Rychlik and Wallin \cite{Maghsood3}, the Viterbi algorithm was used to calculate the number the driving events. However, the Viterbi algorithm requires access to the entire data sequences and thus can not be used for online estimation when the data is not stored. Instead we compute the expected number of events as follows:

%The Viterbi algorithm determines the most likely sequence $\hat{X}_{t}$ of hidden states (events), which maximizes the conditional probability of the observation sequence for given parameters $\Theta$:
%\begin{align*}
% \hat{X}_{1:n} = \argmax_{X_{1:n}}&P(Y_{1},...,Y_{n}| X_{1:n};\Theta).
%\end{align*}

Suppose that at each time $t$, the Markov chain $\{Z_t\}$ with transition matrix $\mv{Q}_t$  by solving equation $(\mv{Q}_t-I)\mv{\pi_t}=\mv{0}$, one gets  the stationary distribution of  $\mv{Q}_t$ . If the data comes from a stationary distribution then $\mv{\pi_t}$ would be the stationary distribution of $\{Z_t\}$. If the data is not stationary one could estimate the stationary distribution by taking the average, over time, of $\mv{\pi_t}$.   By the same reasoning we estimate the expected number of $i^{th}$ event up to time $T$ as
\begin{align}\label{IntEvents}
\eta_i(T)=E[\sum^{T}_{t=1} \xi_i(t)]=\sum^{T}_{t=1} \sum_{j\neq i}\pi_{t,j} q_t(j,i),
\end{align}
where $\xi_i(t)=\sum_{j\neq i} I (Z_{t}=j,Z_{t+1}=i)$.

The above formula works if we substitute $\mv{Q}_t$ with the online estimate $\hat{\mv{Q}}_t$ for each $t$. Then, one can compute  and update the number of events based on each new observation.

\subsection{HMMs with Laplace distribution}
As mentioned in the introduction, we set the conditional distribution of $Y_t$ given $Z_t$, denoted by $g_{\theta}(i,y_t)$, to be a generalized asymmetric Laplace distribution (GAL), see \cite{kotz}. The GAL distribution is a flexible distribution with four parameters: $\mv{\delta}-$ location vector, $\mv{\mu}-$ shift vector, $\nu>0-$ shape parameter, and $\mv{\Sigma}- $ scaling matrix and denoted by $GAL(\mv{\delta},\mv{\mu}, \nu, \mv{\Sigma})$. The probability density function (pdf) of a $GAL(\mv{\delta},\mv{\mu}, \nu, \mv{\Sigma})$ distribution is
\begin{eqnarray}
g(\mv{y}) &=& \frac{1}{\Gamma(1/\nu) \sqrt{2\pi}} \left( \frac{\sqrt{(\mv{y}-\mv{\delta})^T\mv{\Sigma}^{-1}(\mv{y}-\mv{\delta}) }}{c_2}\right)^{\frac{1/\nu-d/2}{2}} e^{(\mv{y}-\mv{\delta})\Sigma^{-1}\mv{\mu}}\nonumber\\
&\,&\qquad\qquad\qquad\qquad\qquad K_{1/\nu-d/2} \left( c_2\sqrt{(\mv{y}-\mv{\delta})^T\mv{\Sigma}^{-1}(\mv{y}-\mv{\delta}) }\right),\nonumber
\end{eqnarray}
%$$
%f(x) = \frac{1}{\Gamma(\lambda) \sqrt{2\pi}} \left( \frac{|x-\delta|}{c_2}\right)^{\nu-0.5} e^{\frac{(x-\delta)\mu}{\sigma^2}} K_{\nu-0.5} \left( c_2|x-\delta|\right),
%$$
where $d$ is the dimension of $\mv{Y}$, $c_2 = \sqrt{2+\mv{\mu}^T\mv{\Sigma}^{-1}\mv{\mu}}$ and $K_{1/\nu - d/2}(.)$ is the modified Bessel function of the second kind. The normal mean variance mixture representation can give an intiutive feel of the distribution. That is a random variable $\mv{Y}$ having GAL distribution and the following equality works:
$$
\mv{Y} \overset{d}{=}  \mv{\delta} + \Gamma \mv{\mu} + \sqrt{\Gamma} \mv{\Sigma}^{1/2} \mv{Z},
$$
where $\Gamma$ is a Gamma distributed random variable with shape $1/\nu$ and scale one, and $\mv{Z}$ is a vector of $d$ independent standard normal random variable. For more details see \cite{Barndorff}.

%O. Barndorff-Nielsen, J. Kent, and M. Sørensen. Normal variance-mean mixtures and z distributions. International Statistical Review, 50:145–159, 1982.

%%%%%%%%%%%%%%%%%%%%%%%%%%%%%%%%
\section{Estimation of fatigue damage}
%From a probabilistic point of veiw, the load is defined as a random processes and the fatigue is a deterministic function of this random processes.

Fatigue is a random process of material deterioration caused by variable stresses. For a vehicle, stresses depend on environmental loads, like road roughness, vehicle usage or driver's behavior. 

Often, the rainflow cycles are calculated in order to describe the environmental  loads \cite{kotz}, and the fatigue damage is then approximated by a function of the rainflow cycles. 

Typically, the approximations are done in order to reduce the length of the load signals storing only the events relevant for fatigue. The reduced signal is then used to find the fatigue life of components in a laboratory (or to estimate the fatigue life mathematically). The reduction is mainly done in order to speed up the testing which is very expensive (or simplify calculations). 

In this section, we present a method to approximate the environmental load using driving events. The method is similar to a well-known method in fatigue analysis, the rainflow filter method  \cite{Johannesson2}. We show that one can explicitly calculate the expected damage intensity (which describes the expected life time of a component) online. 

We start with a short introduction to rainflow cycles and expected damage, then show the approximation method that uses the driving event to derive the expected damage.

\subsection{Rainflow counting distribution and the expected damage} \label{sec:damageindex} 
%The most common way to compute the fatigue damage is by means of rainflow cycles. The rainflow cycle count algorithm is one of the most commonly used methods to count the cycles. 
The rainflow cycle count algorithm is one of the most commonly used methods to compute fatigue damage. The method was first proposed by Matsuishi and Endo \cite{Matsuishi}. Here, we  use the definition given by Rychlik \cite{Rychlik1} which is more suitable for statistical analysis of damage index. The rainflow cycles are defined as follows.

\begin{figure}[H] 
\centering
\includegraphics[width=8cm]{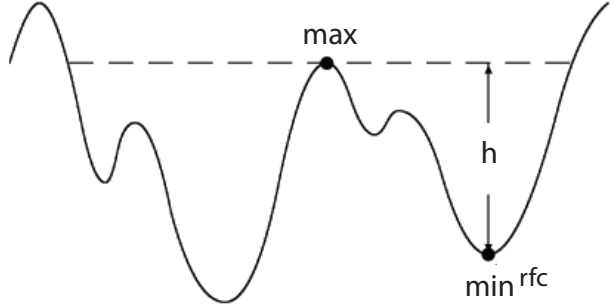}
\caption{The rainflow cycle.}
\label{fig:rfc}
\end{figure}

Assume that a load $L_T$, the processes up to time $T$, has $N$ local maxima. Let $M_i$ denote the height of $i^{th}$ local maximum. Denote $m^{+}_{i}$ ($m^{-}_{i}$) the minimum value in forward (backward) direction from the location of $M_i$ until $L_T$ crosses $M_i$ again. The rainflow minimum, $m^{rfc}_{i}$, is the maximum value of $m^{+}_{i}$ and $m^{-}_{i}$. The pair $(m^{rfc}_{i},M_{i})$ is the $i^{th}$ rainflow pair with the rainflow range $h_{i}(L_T) =M_{i}-m^{rfc}_{i}$. Figure~\ref{fig:rfc} illustrates the definition of the rainflow cycles.\\[0.5cm]
By using the rainflow cycles found in $L_T$, the fatigue damage can be defined by means of Palmgren-Miner (PM) rule~\cite{Palmgren}, \cite{Miner},
\begin{equation}\label{Damind}
D_\beta(L_T) = \alpha \sum_{i=1}^N h_{i}(L_T)^\beta,
\end{equation}
where $\alpha,\beta$ are material dependent constants.
The parameter $\alpha^{-1}$ is equal to the predicted number of cycles with range one leading to fatigue failure (throughout the article it is assumed that $\alpha$ equals one). Various choices of the damage exponent $\beta$ can be considered, like $\beta=3$ which is the standard value for the crack growth process or $\beta=5$ which is often used when a fatigue process is dominated by the crack initiation phase.

A more convenient representation, from computational viewpoint, of damage is:
\begin{align} \label{eq:D41}
D_\beta(L_T)=\beta(\beta-1)\int^{+\infty}_{-\infty}\int^{v}_{-\infty}\,(v-u)^{\beta-2}N^{osc}(u,v)\,du\,dv,
\end{align}
where $N^{osc}(u,v)$ is the number of interval ($[u,v]$) upcrossing by a load, see \cite{Rychlik2} for details.

Since $L_T$ is a random process, one uses the expected damage as a tool to describe damage. The damage intensity of a process is
%The average growth of the expected damage is:
\begin{equation}\label{dint}
d_\beta=\lim_{T\rightarrow\infty}\frac{1}{T} E[D_\beta(L_T)].
\end{equation}
Finally, using Eq.~(\ref{eq:D41}), we get that
\begin{align}
d_\beta=\beta(\beta-1)\int^{+\infty}_{-\infty}\int^v_{-\infty} (v-u)^{\beta-2}\mu^{osc}(u,v)\,du\,dv,
\label{eq:D7}
\end{align}
%To compute this quantity one replaces $N^{osc}(u,v)$ in (\ref{eq:D41}) with 
where
\begin{equation}\label{int_cross}
\mu^{osc}(u,v)=\lim_{T\rightarrow\infty}\frac{E\left[N^{osc}(u,v)\right]}{T}.
\end{equation} 
which is called the intensity of interval up-crossings.
\subsection{Reduced load and expected damage given driving events}
%As mentioned before, the reduced loads are often used to verify the fatigue life of a component. There are several methods to reduce the load. The popular one is the so called rainflow filtering which removes local extremes located at the top and bottom of rainflow cycles with amplitude $h$ below a suitable chosen threshold. Note that the rainflow filtering is just a post-processing of measured load to obtain a shorter load sequences. In this section, we present an alternative method to reduce the load applying the identified driving events. 

 In general the lateral loads are not available and will vary between vehicles. The reduced load, we propose below, is constructed using estimated frequencies of driving events from the HMM, and the distributions of extreme loads associated with driving events, which can be  measured on testing grounds or in laboratories.

 We now describe how to construct a reduced load from the driving events left turn, $LT$, and right turn, $RT$ (the method could of course be generalized to other driving events);  these events are known to cause the majority of the damage for steering components.
 % Let $Z^{*}_i$ represents the $i^{th}$ turn, occurring in the time interval $[t_{i,start},t_{i,stop}]$, and be equal one if the turn is left, and two if the turn is right. Note that the driving states $Z_t$ of HMM is defined on $t=0, \dots, T$, while $Z^{*}_i$ is defined on  intervals $[t_{i,start},t_{i,stop}]$ for $i=0, \dots, N$. That is, the value of $Z_t$ at time $t$ represent one of three driving states right turn, left turn or straight forward; whereas, an sequence of  $Z_{t:t+k}$ which are all equal, to the same driving state, is a driving event. 
Let $\{Z_t\}_{t=0}^T$ be the hidden processes in a HMM, with three possible driving states right turn, left turn or straight forward, at time $t$. Here, we define $Z^{*}_i$  as the driving event representing the $i^{th}$ turn, occurring in the time interval $[t_{i,start},t_{i,stop}]$, and is equal one if the turn is left, and two if the turn is right.  The relation between the two sequences $\{Z^{*}_i\}_{i=0}^N$ and $\{Z_t\}_{t=0}^T$ is that the event $\{Z^{*}_i=1\}(\mbox{ or }\{Z^{*}_i=2\})$ is equivalent to that $Z_{t_{i,start}},..., Z_{t_{i,stop}}$ are all equal to, the same driving state, left turn (or right turn).

\begin{figure}[H] 
\centering
\includegraphics[width=11cm]{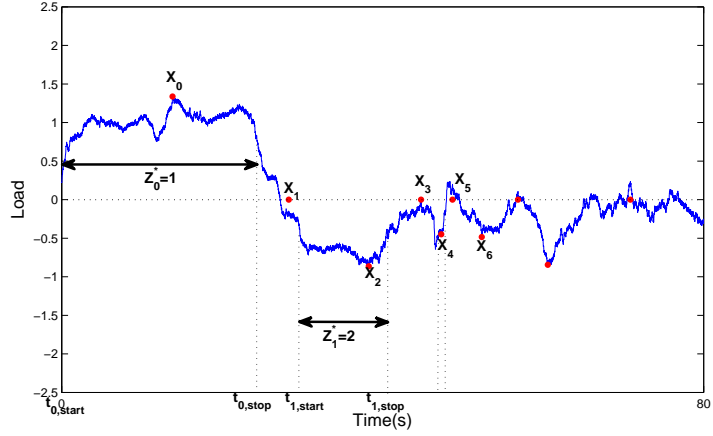}
\caption{Reduced load represented by dots where the observed load is represented by the irregular solid line.}
\label{fig:MinMax}
\end{figure}

Now to create the reduced load, from the sequence driving events, assume that $M_i$ and $m_i$ are the $i^{th}$ maximum and minimum load during a turn, that is
\begin{align}
M_i=\displaystyle \max_{t \in I_i}L_t,\qquad m_i=\displaystyle\min_{t \in I_i}L_t,
\label{eq:Mm}
\end{align}
where $I_i= [t_{i,start},t_{i,stop}]$ represents the start and stop points of $i^{th}$ turn. The reduced load $\left\{X_i\right\}^{N}_{i=0}$ is defined as follows 
\begin{align}
X_i =
\begin{cases}
0, &\mbox{if $i$ is odd integer},\\
M_{i/2}, &\mbox{if } Z^{*}_i=1, \mbox{$i$ is even integer},\\
m_{i/2}, &\mbox{if } Z^{*}_i=2, \mbox{$i$ is even integer}.
\end{cases}
\label{eq:RV}
\end{align}
Here the zeros are inputed since between each left and right turn event there must be a straight forward event.
Figure~\ref{fig:MinMax} illustrates a lateral load and the corresponding reduced load.

\noindent
To compute the damage intensity $d_\beta$, per driving event, one needs the interval up-crossing intensity $\mu^{osc}(u,v)$ of $\{Z^{*}_i\}_{i=0}^N$. Assuming that both $\{M_i\}_{i=0}^N$ and $\{m_i\}_{i=0}^N$ are sequences of iid r.v, and that  the transition matrix $\mv{P}$ of  $Z^*$ is known (it can be derived from transition matrix $\mv{Q}$ in the HMM, see \ref{app:A}), one gets the closed form solution
\begin{align}
\mu^{osc}(u,v)=\frac{1}{2}
\begin{cases}
\pi'_{2}P(m_1<u), &\mbox{$u<v<0$}, \\
\pi'_2\,P(m_1<u)\,p_{2}(u,v), &\mbox{$u\leq 0 \leq v$}, \\
\pi'_{1}P(M_1>v), &\mbox{$0<u<v$}.
\end{cases}
\label{eq:mu1}
\end{align}
Here $\pi'=(\pi'_1,\pi'_2)$ is the stationary distribution of the $\mv{P}$ and $p_{2}(u,v)$ can be derived from the systems
\begin{align}
p_{j}(u,v)=&\, p(j,1)P(M_1>v) + P(M_1\le v)\,p(j,1)\,p_1(u,v)\nonumber\\
& + P(m_1\ge u)\,p(j,2)\,p_2(u,v), \, j=1,2.
\label{eq:p2}
\end{align}
 For more details see \cite{Maghsood3}.

\section{Examples}
We evaluate the proposed algorithm with simulated and measured data sets. We consider the steering events occurring when the vehicle is driving at a speed higher than 10 km/h, e.g.\ when driving in curves. We estimate the number of left and right turns for a costumer. We further investigate the damage caused by steering events and compute the expected damage using the online estimation of transition matrix. 

In our simulation study, a training set is used to estimate the parameters of the model which contains all steering events. We also use the simulation study to show the effects of different values of forgetting factor $\gamma$.

Finally, we use the measured data which is dedicated field measurements from a Volvo Truck. The measured signals come from the CAN (Controller Area Network) bus data, which is a systematic data acquisition and contains customer data. 

\subsection{Simulation study}
We want to imitate a real journey during different road environments, such as city streets and highways. This is done by first generating a  sequence of steering states using a Markov chain. We consider three states right turn (RT), left turn (LT) and straight forward (SF). We set these events as three hidden states and construct the HMM based on them as follows: We assume that the probabilities of going from a right turn to a left turn and vice versa are small and most often we will have straight forward after a right or a left turn. It has been also assumed that the average duration of straight forward during a city road is less than highway. Two different transition matrices $\mv Q_{city}$ and $\mv Q_{highway}$ have been considered for city and highway respectively:

\begin{math}
\mv Q_{city}=
\bordermatrix{&\textrm{RT}&\textrm{SF}&\textrm{LT} \cr
\textrm{RT} & 0.85 &   0.1  &  0.05\cr
\textrm{SF} & 0.025  &  0.95 &   0.025\cr
\textrm{LT} & 0.05 &  0.1 &   0.85\cr
},	
\end{math} 
\begin{math}
\mv Q_{highway}=
\bordermatrix{&\textrm{RT}&\textrm{SF}&\textrm{LT} \cr
\textrm{RT} & 0.90 &   0.08  &  0.02\cr
\textrm{SF} & 0.005 &  0.99 &   0.005\cr
\textrm{LT} & 0.02 &  0.08&   0.90\cr
}.	
\end{math}\\

\begin{figure}[H]
\hspace{-1.2cm}
\begin{tabular}{ccc }
(a) & (b)& (c) \\
\includegraphics[width=4.2cm]{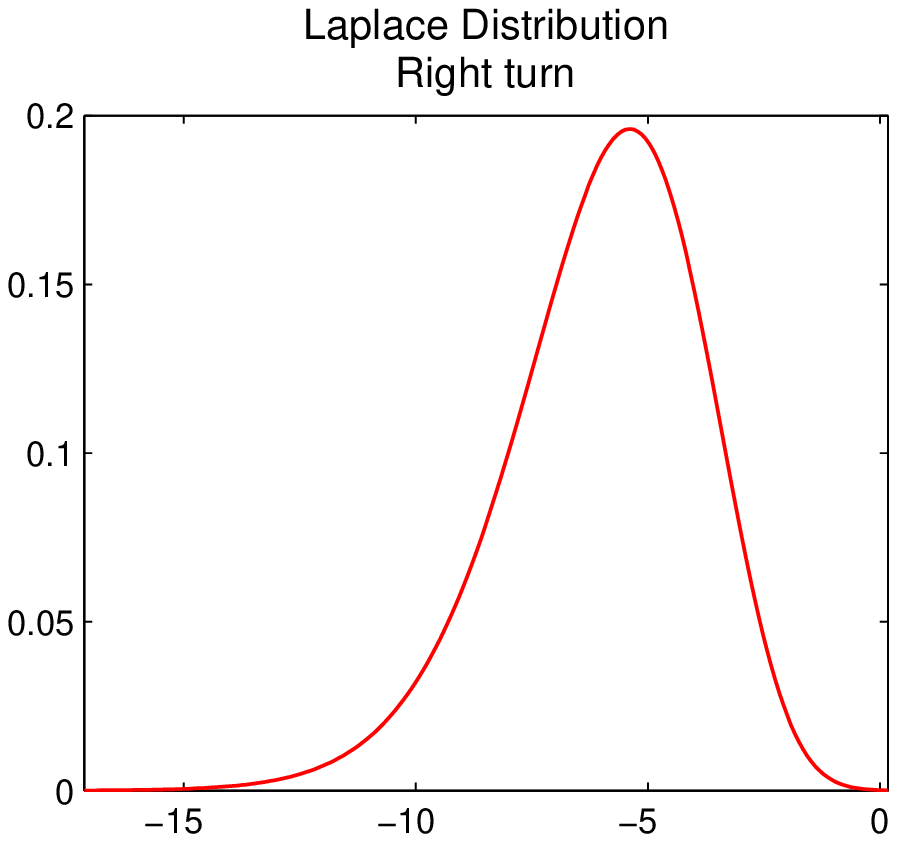} &\includegraphics[width=4.2cm]{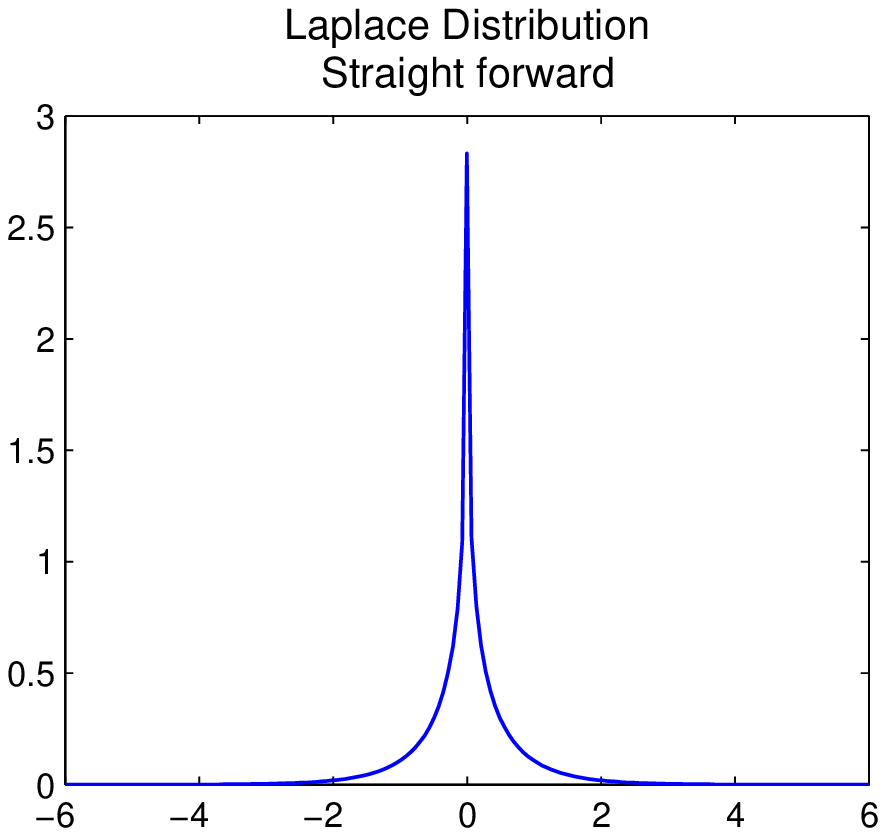}&\includegraphics[width=4.2cm]{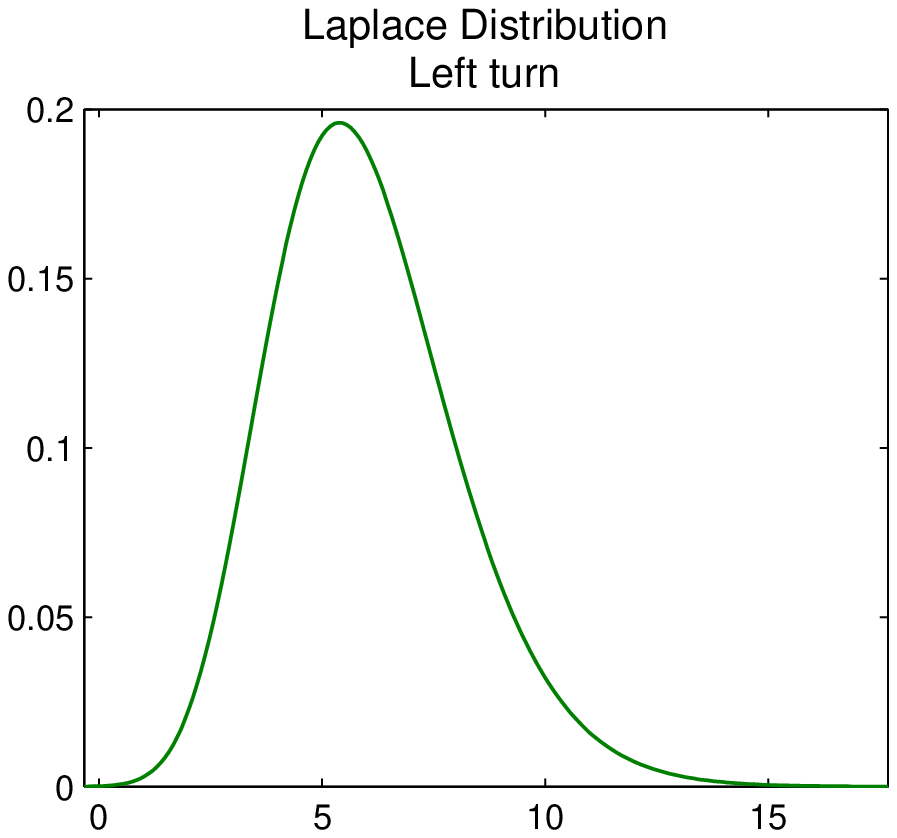}
 \end{tabular}
\caption{(a), (b) and (c) represent the Laplace distributions fitted on lateral acceleration values for right turns, straight forward and left turns respectively.}
\label{fig:dislap}
\end{figure}

\begin{figure}[H] 
\centering
\includegraphics[width=11.5cm]{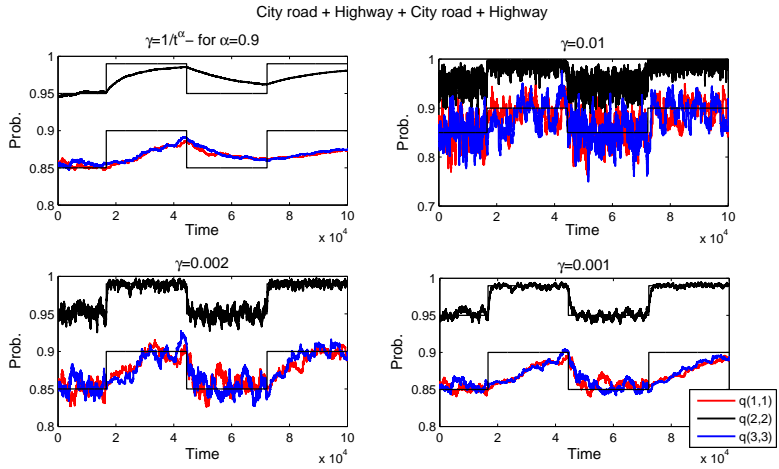}
\caption{Diagonal elements of online estimated transition matrix, simulated signal from City road+Highway+City road+Highway, with four different values of  $\gamma$. Straight thick black lines show the diagonal elements of true transition matrices $\mv Q_{city}$ and $\mv Q_{highway}$. }
\label{fig:EstTra1}
\end{figure}

%\noindent
%In this case, the average duration of each event is:
%\begin{itemize}
%\item[$-$]  Mean time length for SF events in city road: 10s,
%\item[$-$]  Mean time length for SF events in highway: 50s,
%\item[$-$] Mean time length for curves in city road: 3s,
%\item[$-$] Mean time length for curves in highway: 5s.
%%\item[$-$] Mean time length for curves in both environments: 5s.
%\end{itemize}
Second, we use Laplace distribution to simulate the lateral acceleration signal, $Y_t$. The Laplace parameters $(\delta,\mu, \nu, \Sigma)$ for each state are set as follows:
\begin{itemize}
\item $\delta_{RT}=-\delta_{LT}=-1,\ \delta_{SF}=0$,
\item $\mu_{RT}=-\mu_{LT}=-0.5,\ \mu_{SF}=0$,
\item $\nu_{RT}=\nu_{LT}=10,\ \nu_{SF}=0.5$,
\item $\Sigma_{RT}=\Sigma_{LT}=0.2,\ \Sigma_{SF}=1$.
\end{itemize}
The fitted distributions for lateral acceleration values within each state are shown in Figure~\ref{fig:dislap}.
%\textcolor{red}{\mn{RM: From Par comments: Under which assumptions?}}
We compare four different values of $\gamma_t$ for the estimation of the transition matrix. First, we set $\gamma_t=1/t^{\alpha}$ where $\alpha=0.9$. This value of forgetting factor satisfies the convergence conditions given by Capp\'e \cite{Cappe1}. Second we use three different values of fixed $\gamma$, $0.01$, $0.002$ and $0.001$--corresponding to $R=0.9$ and $K=200, 1000$ and $2400$ (which corresponds to a duration $2$ min, $10$ min,  and $20$ min) in Eq.~(\ref{gam}).  Figure~\ref{fig:EstTra1} shows the estimated diagonal elements of the transition matrices for one simulated signal. The simulated signal represents a journey on a city road, a highway and then back to a city road and again highway over $10^5$ seconds,  where the sampling period is $1/2$ seconds. The straight thick black lines show the diagonal elements of true transition matrices $\mv Q_{city}$ and $\mv Q_{highway}$. 
%we lost history about the curves which will happen more frequently during a city road. Finally, we used the small and fixed value $\gamma_t=0.001$ to gain more information from the past. This value of forgetting factor makes estimations more adapted to the environment.

\vspace{-3mm}
In Figure~\ref{fig:EstTra1}, one can see that the online algorithm with variable $\gamma$ can not follow the changes of the parameters well and that the adaption diminishes over time, as is to be expected. The fixed forgetting factor, however, seems to adapt well to the chaining environment. 

\subsubsection*{Expected number of events}
Here, we compute the expected number of turns. We simulate independently hundred signals in order to investigate the accuracy of the online algorithm with different forgetting factors $\gamma$. In that case, we choose as before four different values of forgetting factors, which the fixed values correspond to the weight $R=0.9$ given by the $K=200, 1000$ and $2400$ latest observations in Eq.~$(\ref{gam})$.

We perform 100 simulations and estimate the intensities of occurrences of turns by Eq.~$(\ref{IntEvents})$:
\begin{align}
&\eta_{LT}=\sum_{t=1}^{T}(\pi_{t,2} \hat{q}_t(2,3)+\pi_{t,1} \hat{q}_t(1,3)),\label{countEvents_L}\\
&\eta_{RT}=\sum_{t=1}^{T}(\pi_{t,2} \hat{q}_t(2,1)+\pi_{t,3} \hat{q}_t(3,1)).\label{countEvents_R}
\end{align}
In order to validate the results, we compute an error rate which is the difference between the estimated and observed number of turns in each simulation.  The expected number of turns from the model (using $\mv Q_{city}$ and $\mv Q_{highway}$) are $\eta_{LT}=\eta_{RT}=2840$. The average number of observed left and right turns are $n_{LT}=2834$ and $n_{RT}=2836$, respectively. The average and the standard deviations of errors for 100 simulations are computed. The results  are presented in Table~\ref{tab:DetectionOnline1}. According to the average error, the forgetting factor $\gamma_t=0.002$ performs the best. However there is, surprisingly, only a small difference between all the fixed forgetting factors.
%It can be seen that the Viterbi algorithm underestimates the number of turns.
\begin{table}[H] 
\caption{The expected number of turns estimated by online algorithm and Eqs.~$(\ref{countEvents_L}), (\ref{countEvents_R})$. The errors are the average of the differences between the estimated and observed number of turns.} 
\label{tab:DetectionOnline1}
%\tiny
\footnotesize 
\begin{center}
\begin{tabular}{|l | c  c | c c |  c c |  c c |}
\hline 
\multicolumn{9}{|c |}{Online algorithm}\\  \cline{1-9}

$\gamma_{t}$& \multicolumn{2}{c |}{ $1/t^{0.9}$}& \multicolumn{2}{c |}{ $0.01$} & \multicolumn{2}{c |}{ $0.002$}& \multicolumn{2}{c | }{ $0.001$} \\
 \hline 
Turns & $\eta_{LT}$ & $\eta_{RT}$ & $\eta_{LT}$ & $\eta_{RT}$  & $\eta_{LT}$ & $\eta_{RT}$ & $ \eta_{LT}$  & $\eta_{RT}$ \\
 \hline
Mean\  Est. &  3236 &   3241 &   2928  &  2932 &   2882  &  2886  &  2920  & 2924\\
 \hline
Mean\ Error & 402.48 &  405.30 &  94.40 &  96.46  & 48.45  & 49.93 &  86.84 &  88.68 \\
 \hline
Std\ Error &  28.45 &  33.78  & 15.41 &  15.79  & 16.61  &17.77   & 20.65 &  21.43 \\
 \hline
\end{tabular}
\end{center} 
\end{table}
% $n_{LT}=2830$ and $n_{RT}=2832$
%Viterbi algorithm in average are $\eta_{LT}=2615$ and $\eta_{RT}=2800$.
%Mean\  Est. &  3335   & 3306 &   3239  &  3107  &  3058  & 3006  &  2935  & 2915 \\
% \hline
%Mean\ Error & 504.58 & 474.67 & 408.73  & 275.25 &  227.65 & 174.57  & 104.44  &  83.17 \\
% \hline
%Std\ Error &  56.45 &  49.20 &   75.82&   63.82 &   50.11 &   39.68&   41.67  &  33.67  \\
% \hline
	
In our previous work, an HMM combined with a Viterbi algorithm~\cite{Viterbi} has been used to identify the driving events. The Viterbi algorithm gives a reconstructed sequence of events which maximizes the conditional probability of the observation sequence. In that approach, all data has to be used to estimate the driving events and is thus not suitable to on-board usage in a vehicle. However, in order to compare the previously proposed approach with the online estimation and to evaluate the frequencies of driving events, we also compute the number of turns by the Viterbi algorithm for each simulation. The counted number of turns from the Viterbi algorithm are on average $\eta_{LT}=2923$ and $\eta_{RT}=2925$. One can see that the Viterbi algorithm overestimates the number of turns.

\subsubsection*{Damage investigation}
\begin{figure}[H] 
\centering
\includegraphics[width=12cm]{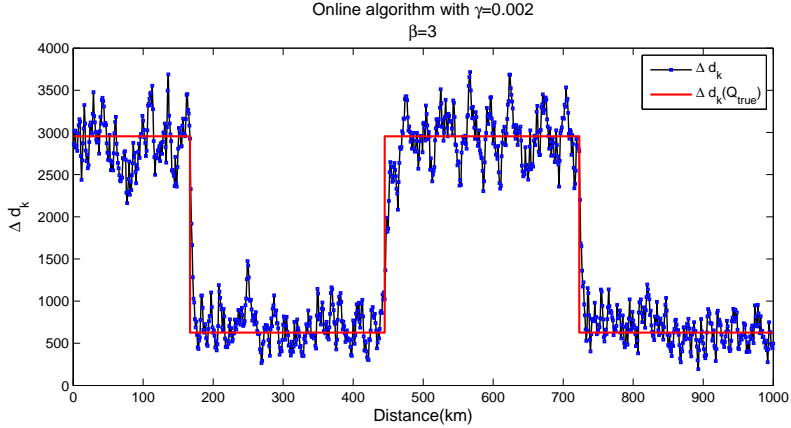}
\caption{Damage intensity per km according to the online estimation of transition matrix with $\gamma=0.002$. The upper plot shows the results for damage exponent $\beta=3$. The straight thick red line shows $\Delta d_k(\mv Q_{true})$ which is the damage intensity computed using model transition matrices $\mv Q_{city}$ and $\mv Q_{highway}$ for city and highway, respectively.}
\label{fig:simdam2}
\end{figure}

In this section we compute the damage intensity based on online estimation of transition matrix per kilometer. We use one of the simulated lateral acceleration signals in order to calculate the damage. The speed of the vehicle is considered 50 kilometers per hour and the mileage is 1000 km (for a sampling period of $1/2$ seconds). We split the signal into 1000 equally sized frames. For each frame, the expected number of turns are computed by $\Delta \eta_k=\eta_k-\eta_{k-1}$ where $\eta_k$ is the estimated number of turns occurring up to $k^{th}$ frame.  The expected damage based on turns for each frame is calculated by:
\begin{align*}
\Delta d_k=\Delta \eta_kd_k,
\end{align*}
where $d_k$ is the expected damage per turn and calculated by means of  Eqs.~(\ref{eq:D7}) and~(\ref{eq:mu1}). The empirical distribution of $M_i$ and $m_i$ are used to calculate the intensity of interval crossings $\mu^{osc}(u,v)$. We use the online estimation of transition matrix $\mv Q$ with $\gamma=0.002$ to estimate the transition matrix $\mv P$ by using Eqs.~(\ref{eq:p11}) and~(\ref{eq:p22}), see \ref{app:A}. 
%The Rayleigh distributions which have been fitted to positive and negative values of the reduced load are
%\begin{equation*}
%%P(M_1>v)=e^{-\frac{1}{2}\left(\frac{v}{3.6}\right)^2},\, v\ge 0, \qquad P(m_1<u)=e^{-\frac{1}{2}\left(\frac{u}{3}\right)^2},\, u\le 0.
%P(M_1>v)=P(m_1<u)=e^{-\frac{1}{2}\left(\frac{v}{6.2}\right)^2},\, v\ge 0,\ u\le 0.
%\end{equation*}
The result for damage exponent $\beta=3$ is shown in Figure~\ref{fig:simdam2}.\  The straight thick red line shows $\Delta d_k(\mv Q_{true})$ which is the damage intensity computed using the model transition matrices $\mv Q_{city}$ and $\mv Q_{highway}$ for city and highway respectively. We can observe the change in damage between highway and city road. As might be expected the damage intensities (per km) estimated for the city are higher than for highway, since the number of turns occurring in a city road are larger than on a highway.

Further, the expected damage from the model (theoretical damage) is compared with the total damage and the damage calculated from the reduced load. One can see that the expected damage for the whole signal -- based on online estimation of transition matrix-- is equal to $\sum ^{1000}_{k=1}\Delta d_k$. The total damage is calculated from the lateral acceleration signal using the rainflow method. The damage evaluated for the load (lateral acceleration), reduced load and the expected damage is compared in Table~\ref{tab:Dam1}. The numerical integration in~(\ref{eq:D7}) as well as the rainflow cycle counting has been done using the WAFO (Wave Analysis for Fatigue and Oceanography) toolbox, see \cite{Brodtkorb2000.C1,WAFO2011.M1}.
\begin{table}[H] 
\caption{Comparison of damage computed for the simulated load, the corresponding reduced load and the expected damage.} 
\label{tab:Dam1}
\begin{center}
\begin{tabular}{|l| c|  c|c| }
 \hline 
\multirow{ 2}{*}{Damage} & \multirow{ 2}{*}{Total} & \multirow{ 2}{*}{Reduced load}  & Expected\\ 

 & & & Online\ with $\gamma=0.002$\\
 \hline
 $\beta=3$ & $1.88\cdot 10^6$  & $1.68\cdot 10^6$ & $1.68\cdot 10^6$ \\
 \hline
 $\beta=5$ & $1.77\cdot 10^8$ & $1.72\cdot 10^8$  &  $1.67\cdot 10^8$ \\
 \hline
	\end{tabular}
	\end{center} 
	\end{table}
Figure~\ref{fig:simdam2} and Table~\ref{tab:Dam1} demonstrate high accuracy of the proposed approach to estimate the expected damage for the studied load. Obviously this load is a realistic mathematical model of a real load. In the next section we will apply our method to estimate the steering events and compute the damage for a measured load on a VOLVO truck.

\subsection{On-board logging data from Volvo}
To evaluate the method on a real data set, we study field measurements coming from a Volvo Truck. We use the measured lateral acceleration signal from the CAN (Controller Area Network) bus data. 

We fit the Laplace distribution for the lateral acceleration within each steering state. To estimate the Laplace distribution parameters considered, we need a training set which contains all history about the curves. We detect the events manually by looking at video recordings from the truck cabin to see what happened during the driving. The manual detections are not completely correct because of the visual errors and the low quality of videos used for the manual detection. 

%Here, we use a simple naive algorithm which automatically identifies the driving events. The method detects the curves by using predefined thresholds for lateral acceleration signals. If the absolute value of lateral acceleration is larger than $0.2 \ \textrm{m/s}^2$ for more than 1.5 seconds, then the algorithm will detect the event as a turning event. 

\begin{figure}[H] 
\centering
\includegraphics[width=11cm]{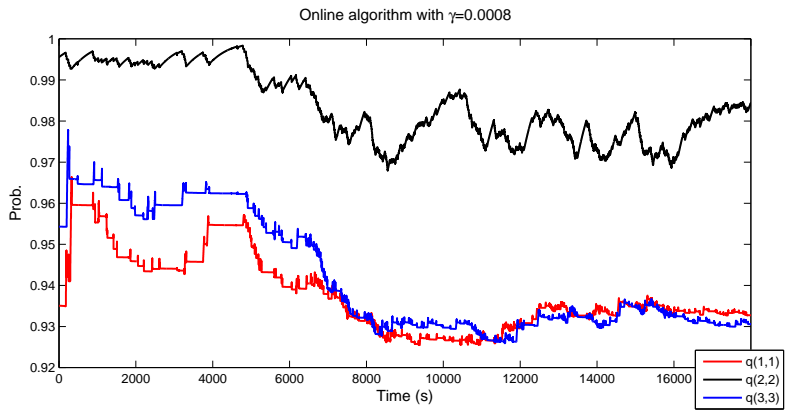}
\caption{Diagonal elements from online estimation of transition matrix with $\gamma_t=0.0008$ for measured data}
\label{fig:meas1}
\end{figure}

%Road conditions can be divided into different classes and a customer will drive different distances in each road class. Hence, we have tried to find the value of forgetting factor which is more adapted to the environment. Both online and EM algorithms have been used to count the number of steering events.  
The online algorithms are used to count the number of left and right turns. Figure~\ref{fig:meas1} shows the estimation results using online algorithm with $\gamma_t=0.0008\, (R=0.8, K = 2000)$ for the measured signal. It is interesting to note that there is a sudden change in the driving environment after around 5000 sec.
The expected number of left and right turns computed by online algorithm are $\eta_{LT}=228$ and $\eta_{RT}=241$ respectively.

%Table~\ref{tab:DetectionOnlineMeas} shows the expected number of curves computed using online algorithm which are compared with the detected number of curves computed using EM and Viterbi algorithms. 
%\begin{table}[H] 
%\caption{The expected number of curves computed using online algorithm compared with the detected number of curves computed using EM and Viterbi algorithms.} 
%\label{tab:DetectionOnlineMeas}
%\begin{center}
%\begin{tabular}{|l | c  c | c c | c c | c c|}
%\hline 
%\multicolumn{7}{|c |}{Online}& \multicolumn{2}{c| }{\multirow{ 2}{*}{EM}}\\  \cline{1-7}
%
%Forgetting\ factor& \multicolumn{2}{c |}{ $\gamma_{t}=1/t^{0.9}$}& \multicolumn{2}{c |}{ $\gamma_{t}=0.001$}& \multicolumn{2}{c | }{ $\gamma_{t}=0.0005$} & \multicolumn{2}{c | }{}\\
% \hline 
%Number\ of\ events& $n_{RT}$ & $n_{LT}$ & $n_{RT}$ & $n_{LT}$ & $n_{RT}$ & $n_{LT}$ & $n_{RT}$ & $n_{LT}$\\
% \hline
%Estimated & 206 & 188& 242 & 231 &  237 & 225 &211 & 210\\
%\hline
%	\end{tabular}
%	\end{center} 
%	\end{table}
\subsubsection*{Damage investigation}
\begin{figure}[H] 
\centering
\includegraphics[width=12cm]{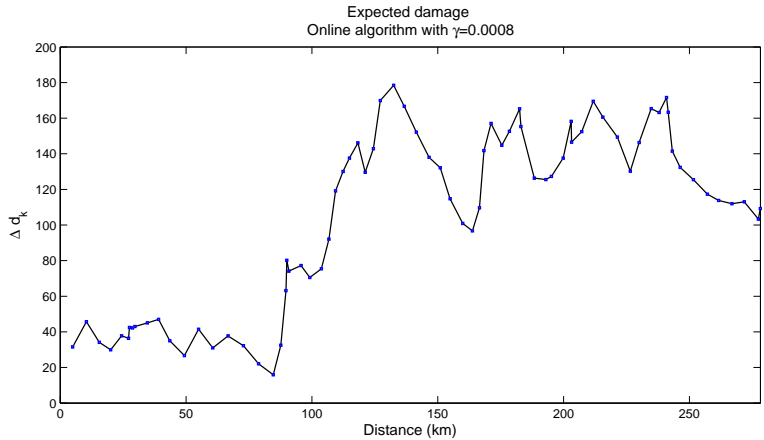}
\caption{Damage intensity with damage exponent $\beta=3$ regarding mileage. The online estimation of transition matrix with $\gamma=0.0008$ has been used to estimate the expected damage.}
\label{fig:measdam2}
\end{figure}

Here, we compute the damage intensity based on the model. In order to do that we split data into the frames containing $250$ seconds (approximately 4-5 km) of measurement and we compute the distance based on the average speed in each frame. Figure~\ref{fig:measdam2} shows the expected damage based on turns computed by $\Delta d_k=\Delta \eta_kd_k$ where $\Delta \eta_k=\eta_k-\eta_{k-1}$ and $n_k$ is the estimated number of turns occur over $k$ kilometers.  Here, the results are based on the damage exponent $\beta=3$.

The total expected damage using online estimation of transition matrix can be computed by $\sum_{k=1}\Delta d_k$. The damage evaluated for the load (lateral acceleration), reduced load and the expected damage is compared in Table~\ref{tab:Dam2}. 
The Rayleigh distributions which have been fitted to positive and negative values of the reduced load are
\begin{equation*}
P(M_1>v)=e^{-\frac{1}{2}\left(\frac{v}{2.2}\right)^2},\, v\ge 0, \qquad P(m_1<u)=e^{-\frac{1}{2}\left(\frac{u}{2.3}\right)^2},\, u\le 0.
\end{equation*}
\begin{table}[H] 
\caption{Comparison of damage values computed from the measured load, the corresponding reduced load and the expected damage.} 
\label{tab:Dam2}
\begin{center}
\begin{tabular}{|l| c|  c|c|}
 \hline 
\multirow{ 2}{*}{Damage} & \multirow{ 2}{*}{Total} & \multirow{ 2}{*}{Reduced load}  & Expected\\ 

 & & & Online\ with $\gamma=0.0008$ \\
 \hline
 $\beta=3$ & $8.1\cdot 10^3$  & $7.4\cdot 10^3$ & $7.7\cdot 10^3$ \\
 \hline
 $\beta=5$ & $1.5\cdot 10^5$ & $1.5\cdot 10^5$ &  $1.9\cdot 10^5$ \\
 \hline
	\end{tabular}
	\end{center} 
	\end{table}

\begin{figure}[H] 
\centering
\includegraphics[width=12cm]{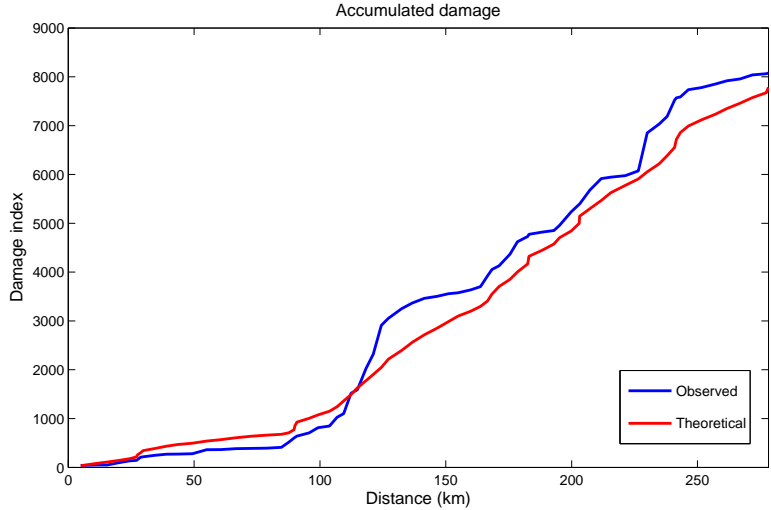}
\caption{The theoretical  and observed accumulated damage processes for damage exponent $\beta=3$. The online estimation of transition matrix with $\gamma=0.0008$ has been used to estimate the expected damage.}
\label{fig:measdam3}
\end{figure}

We also compare the damage accumulation process from the model, $\sum_{k=1}\Delta d_k$, with the empirical accumulated damage in the signal. The expected damage based on fitted model will be called the theoretical damage. Figure~\ref{fig:measdam3} shows the theoretical and observed accumulated damage processes. It can be seen that the accumulated damage from the model is close to the observed damage and there are two damage rates in both theoretical and observed damage processes.

Results shown in Figure~\ref{fig:measdam3} and Table~\ref{tab:Dam2} demonstrate the accuracy of the proposed methodology for this measured load.

%%%%%%%%%%%%%%%%%%%%%%%%%%%%%%%%%%%%%%%%%%%%%%%%%%%%%%%%%%%%%%%%%%%

\section{Conclusion}
In this article, we have derived a method to estimate the number of driving events for a vehicle using the CAN data through the use of an HMM.  The method uses an online EM algorithm to estimate the parameters of the HMM. The online version has three major advantages over the regular EM algorithm, making it possible to implement the method on-board a vehicle:  the computational complexity of each iteration of the algorithm is $\mathcal{O}(1)$, making it a computationally tractable method; the parameters are estimated without the need to store any data; the formulation of the online algorithm allows for an adaptive parameter estimation method, using a fixed forgetting factor, so that the parameters can adapt over chaining driving environment.

The proposed estimation algorithm was validated using simulated and measured data sets. The results show that the online algorithm works well and can adapt to a chaining environment when the driving conditions are not constant over time. 

\section*{Acknowledgment}
We are thankful to Prof.\ Igor Rychlik and Dr.\ P\"ar Johannesson for their useful ideas and helpful suggestions in this study. We would like to thank Volvo Trucks for supplying the data in this study and to the members in our research group at Volvo for their valuable advice. Finally, we gratefully acknowledge the financial support from VINNOVA. The second author has been supported by the Knut and Alice Wallenberg foundation.
%%%%%%%%%%%%%%%%%%%%%%%%%%%%%%%%
%%%%%%%%%%%%%%%%%%%%%%%%%%%%%%%%%%%%%
\nocite{*}
\bibliographystyle{plain} 
\bibliography{Refpaper4}

\begin{thebibliography}{10}

\bibitem{arvastson2000asymptotic}
L~Arvastson, H~Olsson, and J~Holst.
\newblock Asymptotic bias in parameter estimation of ar-processes using
  recursive least squares with exponential forgetting.
\newblock {\em Scandinavian Journal of Statistics}, 27(1):177--192, 2000.

\bibitem{Barndorff}
O.~Barndorff-Nielsen, J.~Kent, and M.~Sorensen.
\newblock Normal variance-mean mixtures and z distributions.
\newblock {\em International Statistical Review}, 50:145--159, 1982.

\bibitem{}
A.~K. Bengtsson and I.~Rychlik.
\newblock Uncertainty in fatigue life prediction of structures subject to
  gaussian loads.
\newblock {\em Probabilistic Engineering Mechanics}, 2009.

\bibitem{Berndt}
H.~Berndt and K.~Dietmayer.
\newblock Driver intention inference with vehicle onboard sensors.
\newblock In {\em IEEE International Conference on Vehicular Electronics and
  Safety (ICVES)}, pages 102--107, Pune, 11-12 November 2009.

\bibitem{Murakami}
A.~Beste, K.~Dressler, H.~K\"{o}tzle, W.~Kr\"{u}ger, B.~Maier, and J.~Petersen.
\newblock Multiaxial rainflow -- a consequent continuation of {P}rofessor
  {T}atsuo {E}ndo's work.
\newblock In Y.~Murakami, editor, {\em The Rainflow Method in Fatigue}, pages
  31--40. Butterworth-Heinemann, 1992.

\bibitem{bishop}
Bishop and Sherratt.
\newblock A theoretical solution for estimation of rainflow ranges from power
  spectral density data.
\newblock {\em Fatigue Frac Eng Mater Struct}, 13:311--326, 1990.

\bibitem{Bogsjo}
K.~Bogsj\"o, K.~Podgorski, and I.~Rychlik.
\newblock Models for road surface roughness.
\newblock {\em Vehicle System Dynamics}, 50:725--747, 2012.

\bibitem{Brodtkorb2000.C1}
P.~A. Brodtkorb, P.~Johannesson, G.~Lindgren, I.~Rychlik, J.~Ryd\'{e}n, and
  E.~Sj\"{o}.
\newblock {WAFO} -- a {M}atlab toolbox for analysis of random waves and loads.
\newblock In {\em Proceedings of the 10th International Offshore and Polar
  Engineering conference, Seattle}, volume III, pages 343--350, 2000.

\bibitem{Cappe1}
O.~Capp\'e.
\newblock Online {EM} algorithm for hidden {M}arkov models.
\newblock {\em Journal of Computational and Graphical Statistics},
  20:3:728--749, 2011.

\bibitem{Cappe}
O.~Capp\'e, E.~Moulines, and T.~Ryd\'en, editors.
\newblock {\em Inference in Hidden {M}arkov Models}.
\newblock Springer, 2005.

\bibitem{Dempster}
A.~P. Dempster, N.~M. Laird, and D.~B. Rubin.
\newblock Maximum likelihood from incomplete data via {EM} algorithm.
\newblock {\em Journal of the Royal Statistical Society. Series B
  (Methodological)}, 39(1):1--38, 1977.

\bibitem{Frendahl}
M.~Frendahl and I.~Rychlik.
\newblock Rainflow analysis - {M}arkov method.
\newblock {\em Int. J. Fatigue}, 15:265--272, 1993.

\bibitem{F-Schnatter}
Sylvia Fr\"{u}hwirth-Schnatter.
\newblock {\em Finite Mixture and {M}arkov Switching Models}.
\newblock Springer, 2006.

\bibitem{Johannesson1}
P.~Johannesson.
\newblock Rainflow cycles for switching processes with {M}arkov structure.
\newblock {\em Probability in the Engineering and Informational Sciences},
  12:143--175, 1998.

\bibitem{Johannesson}
P.~Johannesson.
\newblock {\em Rainflow Analysis of Switching Markov Loads}.
\newblock PhD thesis, Lund Institute of Technology, 1999.

\bibitem{Johannesson2}
P.~Johannesson and M.~Speckert, editors.
\newblock {\em Guide to Load Analysis for Durability in Vehicle Engineering}.
\newblock Wiley:Chichester, 2013.

\bibitem{Karlsson}
M.~Karlsson.
\newblock {\em Load Modelling for Fatigue Assessment of Vehicles -- a
  Statistical Approach}.
\newblock PhD thesis, Chalmers University of Technology, Sweden, 2007.

\bibitem{kotz}
S.~Kotz, T.~Kozubowski, and K.~Podgorski.
\newblock {\em The {L}aplace distribution and generalizations: a revisit with
  applications to communications, economics, engineering, and finance}.
\newblock Springer Science \& Business Media, 2001.

\bibitem{Krenk}
S~. Krenk and H.~Gluver.
\newblock A markov matrix for fatigue load simulation and rainflow range
  evaluation.
\newblock {\em Struct Saf}, 6:247--258, 1989.

\bibitem{Kvanstrom}
M.~Kvanstr\"om, K.~Podg\'orski, and I.~Rychlik.
\newblock Laplace moving average model for multi-axial responses in fatigue
  analysis of a cultivator.
\newblock {\em Probabilistic Engineering Mechanics}, 34:12--25, 2013.

\bibitem{Lindgren04}
G.~Lindgren and KB. Broberg.
\newblock Cycle range distributions for gaussian processes - exact and
  approximate results.
\newblock {\em Extremes}, 7:69--89, 2004.

\bibitem{ljung1983theory}
Lennart Ljung and Torsten S{\"o}derstr{\"o}m.
\newblock Theory and practice of recursive identification.
\newblock 1983.

\bibitem{Maghsood2}
R.~Maghsood.
\newblock A statistical approach for detecting driving events and evaluating
  their fatigue damage, Lic. Thesis, Chalmers University of Technology, 2014.

\bibitem{Maghsood1}
R.~Maghsood and P.~Johannesson.
\newblock Detection of the curves based on lateral acceleration using hidden
  {M}arkov models.
\newblock {\em Procedia Engineering}, 66:425--434, 2013.

\bibitem{Maghsood3}
R.~Maghsood, I.~Rychlik, and J.~Wallin.
\newblock Modeling extreme loads acting on steering components using driving
  events.
\newblock {\em Probabilistic Engineering Mechanics}, 41:13--20, 2015.

\bibitem{Matsuishi}
M.~Matsuishi and T.~Endo.
\newblock Fatigue of metals subjected to varying stress.
\newblock {\em Japan Society of Mechanical Engineers}, 1968.
\newblock \emph{In Japanese}.

\bibitem{Miner}
M.~A. Miner.
\newblock Cumulative damage in fatigue.
\newblock {\em Journal of Applied Mechanics}, 12:A159--A164, 1945.

\bibitem{Mitrovic1}
D.~Mitrovi\'c.
\newblock {\em Learning Driving Patterns to Support Navigation}.
\newblock PhD thesis, University of Canterbury, New Zealand, 2004.

\bibitem{Mitrovic2}
D.~Mitrovi\'c.
\newblock Reliable method for driving events recognition.
\newblock {\em IEEE Transactions on Intelligent Transportation Systems},
  6(2):198--205, 2005.

\bibitem{Palmgren}
A.~Palmgren.
\newblock Die {L}ebensdauer von {K}ugellagern.
\newblock {\em {Z}eitschrift des {V}ereins {D}eutscher {I}ngenieure},
  68:339--341, 1924.
\newblock \emph{In German}.

\bibitem{Rychlik1}
I.~Rychlik.
\newblock A new definition of the rainflow cycle counting method.
\newblock {\em International Journal of Fatigue}, 9:119--121, 1987.

\bibitem{rychlik88}
I.~Rychlik.
\newblock Rain flow cycle distribution for ergodic load processes.
\newblock {\em SIAM J Appl Math}, 48:662--679, 1988.

\bibitem{Rychlik2}
I.~Rychlik.
\newblock Note on cycle counts in irregular loads.
\newblock {\em Fatigue \& Fracture of Engineering Materials \& Structures},
  16:377--390, 1993.

\bibitem{Rychlik13}
I.~Rychlik.
\newblock Note on modelling of fatigue damage rates for non-{G}aussian
  stresses.
\newblock {\em Fatigue \& Fracture of Engineering Materials \& Structures},
  36:750--759, 2013.

\bibitem{Lin}
I.~Rychlik, G.~Lindgren, and Y.K. Lin.
\newblock {M}arkov based correlations of damages in {G}aussian and
  non-{G}aussian loads.
\newblock {\em Probabilistic Engineering Mechanics}, 10:103--115, 1995.

\bibitem{vahidi2005recursive}
Ardalan Vahidi, Anna Stefanopoulou, and Huei Peng.
\newblock Recursive least squares with forgetting for online estimation of
  vehicle mass and road grade: theory and experiments.
\newblock {\em Vehicle System Dynamics}, 43(1):31--55, 2005.

\bibitem{Viterbi}
A.~J. Viterbi.
\newblock Error bounds for convolutional codes and an asymptotically optimal
  decoding algorithm.
\newblock {\em IEEE Transactions on Information Theory}, IT-13(2):260--269,
  1967.

\bibitem{WAFO2011.M1}
{WAFO Group}.
\newblock {WAFO} -- a {M}atlab toolbox for analysis of random waves and loads,
  tutorial for {WAFO} 2.5.
\newblock Mathematical Statistics, Lund University, 2011.

\bibitem{WAFO2011.Ver2.5}
{WAFO Group}.
\newblock {WAFO -- a Matlab Toolbox for Analysis of Random Waves and Loads},
  {Version 2.5}, {07-Feb-2011}.
\newblock Mathematical Statistics, Lund University, 2011.
\newblock \\{\small Web: \texttt{http://www.maths.lth.se/matstat/wafo/}}
  (Accessed 24 January 2014).

\bibitem{Zeitouni}
O.~Zeitouni and A.~Dembo.
\newblock Exact filters for the estimation of the number of transitions of
  finite-state continuous-time {M}arkov processes.
\newblock {\em IEEE Transactions on Information Theory}, 34(4):890--893, 1988.

\end{thebibliography}
\appendix{
\section{Derivation of transition matrix of driving events.}
\label{app:A}
To construct the sequence $\{Z^{*}_i\}_{i=0}^{N}$, of driving events, let $\{t_k\}_{k=0}^{N}$ be the indices in $\{t: Z_{t_k} \neq Z_{t_k-1} \cap  Z_{t_k}  \neq SF\}$, then 
 \begin{align}
 \label{eq:Zk}
 Z^{*}_k =
 \begin{cases}
 1, &\mbox{if } Z_{t_k}=LT,\\
 2, &\mbox{if } Z_{t_k}=RT.
 \end{cases} 
 \end{align}
 Assume that $Z^*$ has transition matrix $\mv{P}=(p(k,j))$. Note that the hidden process $\{Z_t\}_{t=0}^T$ in HMM has three states $"1"=\textrm{RT}$, $"2"=\textrm{SF}$ and $"3"=\textrm{LT}$. 
One can now derive the transition matrix $\mv{P}$ from the transition matrix of the HMM $\hat{\mv{Q}}$ as follows:
\begin{align}
\hat{p}(1,1)&= \frac{ \hat{q}(3,2)\hat{q}(2,3)}{(1-\hat{q}(2,2))(1-\hat{q}(3,3))},\label{eq:p22} \\
\hat{p}(2,2)&=\frac{ \hat{q}(1,2)\hat{q}(2,1)}{(1-\hat{q}(2,2))(1-\hat{q}(1,1))}.\label{eq:p11}
\end{align}
As proof, we consider for instance the probability of going from $\textrm{LT}$ to $\textrm{RT}$  in $Z_i$ which can be computed as follows:
\begin{align*}
\hat{p}(1,2)&=P(Z^{*}_i=1,Z^{*}_{i+1}=2)\nonumber\\
&=P(Z_{t_{i,start}:t_{i,stop}}=3, Z_{t_{i,stop}+1:t_{i+1,start}-1}=2, Z_{t_{i+1,start}:t_{i+1,stop}}=1)\nonumber\\
&\,\qquad\qquad\qquad\qquad+P(Z_{t_{i,start}:t_{i,stop}}=3, Z_{t_{i+1,start}:t_{i+1,stop}}=1),\nonumber\\
&=  \left(\hat{q}(3,2) \left(1+\hat{q}(2,2)+\hat{q}^2(2,2)+...\right)\hat{q}(2,1)+\hat{q}(3,1)\right)\nonumber\\
&\,\qquad\qquad\qquad\qquad\qquad(1+\hat{q}(3,3)+\hat{q}^2(3,3)+...),\\
&= \left[\frac{\hat{q}(3,2)\hat{q}(2,1)}{1-\hat{q}(2,2)}+\hat{q}(3,1)\right]\frac{1}{1-\hat{q}(3,3)}.
\end{align*}
}
where $Z_{t_{i,start}:t_{i,stop}}$ represents the sequence of consecutive driving states $Z_{t_{i,start}},..., Z_{t_{i,stop}}$.
%\lstinputlisting[language=Matlab]{density_GAL.m}
%%%%%%%%%%%%%%%%%%%%%%%%%%%%%
\end{document}